\documentclass[letter]{aa} 
\usepackage{graphicx}
\usepackage[varg]{txfonts}
\usepackage{lscape}
\usepackage{multicol}
\usepackage{color} 
\usepackage[colorlinks, linkcolor=blue,
urlcolor=blue,citecolor=blue,bookmarks, bookmarksopen,
bookmarksnumbered,pdfauthor={Chelli et al}, breaklinks=true]{hyperref} 
\usepackage{epstopdf}
\usepackage[T1]{fontenc}
\newcommand\T{\rule{0pt}{2.6ex}}       
\newcommand\B{\rule[-1.2ex]{0pt}{0pt}} 
\begin{document} 
   \title{Pseudomagnitude Distances: Application to the Pleiades
     cluster}   \author{A. Chelli\inst{1} \and G. Duvert\inst{2,3}} 
   \institute{Université Côte d'Azur, OCA, CNRS, Lagrange, France.
  \thanks{Correspondence: {\tt Alain.Chelli@oca.eu}}
\and Univ. Grenoble Alpes, IPAG, F-38000 Grenoble, France
\and CNRS, IPAG, F-38000 Grenoble, France }

   \date{Received; accepted;}
 
  \abstract
  {The concept of pseudomagnitude was recently introduced by
    \cite{Chelli16}, to estimate apparent stellar diameters using a
    strictly observational methodology. Pseudomagnitudes are distance
    indicators, which have the remarkable property of being reddening
    free. In this study, we use Hipparcos 
    parallax measurements to compute the mean absolute
    pseudomagnitudes of solar neighbourhood dwarf stars as a function of their spectral
    type. To illustrate the use of absolute pseudomagnitudes, we
    derive the distance moduli of $360$
    Pleiades stars and find that the centroid of their distribution is
    $5.715\pm0.018$,
    corresponding to a distance of $139.0\pm1.2$\,pc. We locate the
    subset of $\sim 50$ Pleiades stars observed by Hipparcos
    at a mean distance of $135.5\pm3.7$\,pc, 
    thus confirming the frequently reported anomaly in the Hipparcos
    measurements of these stars. }



   \maketitle
%

\section{Introduction}
In astrophysics, the calculation of interstellar extinction is a
complex and recurring problem. For many objects, such as those buried
in star-forming regions, unreddening the photometries is a difficult
and demanding task.  In the case of a star, the calculation of
interstellar extinction requires a detailed knowledge of its
luminosity class, spectral type, and intrinsic colors. That is a lot
of parameters, not always available, whose robustness is often
uncertain.This leads to the accumulation of errors, and 
makes it nearly impossible to attempt any massive statistical
analysis. 

We recently introduced the concept of 
pseudomagnitude for the calculation of the apparent size of stars,
thus avoiding to deal with the problem of visual extinction
\citep{Chelli14,Chelli16}.  This has allowed us to compile a catalogue
of $453\,000$ angular diameters, with an accuracy of the order of $1\%$
($2\%$ systematic). Pseudomagnitudes are linear combinations of
magnitudes constructed in such a way as to eliminate interstellar
extinction. They are purely observational quantities that are
unaffected by reddening effects, and can be applied to any type of
object. As in the case of magnitudes, pseudomagnitudes are distance
indicators, and absolute pseudomagnitudes, measured at a distance of
$10$\,pc, are luminosity indicators.

Knowledge of the pseudomagnitudes and absolute pseudomagnitudes of
stars allows their distance to be estimated. 
In the present study, we use the parallax measurements of
Hipparcos~\citep{ESAHIP,2007A&A...474..653V} to calculate 
the mean absolute pseudomagnitude of field
dwarf stars, as a
function of their spectral type. As an example, we use this technique
to determine the centroid of the distance distribution of 360 stars in
the Pleiades cluster.

In section~\ref{sec:pseudomagnitudes}, we explain the concept of
pseudomagnitudes. In section~\ref{sec:absolutepseudomagnitudes}, we use 
distance filtered parallax measurements to calculate the mean absolute
pseudomagnitudes (V,J), (V,H) and (V,Ks) of dwarf stars, and the centroid of the distance
distribution of our Pleiades stars is calculated and discussed in
section~\ref{sec:pleiades}. 
\section{Pseudomagnitudes}
\label{sec:pseudomagnitudes}
We define the pseudomagnitude $pm_{\{i,j\}}$ of an astrophysical
object as follows: 
\begin{equation}
pm_{\{i,j\}} = \frac{c_im_j-c_jm_i}{c_i-c_j}
\label{eq:pseudomagnitude}
\end{equation}
where $m_i$ and $m_j$ are the magnitudes measured in the photometric
bands $i$ and $j$, $c_i$ (resp. $c_j$) is the ratio of the interstellar
extinction coefficients $R_i$ and $R_v$ between band 
$i$ and the visible band. We note that when one of the coefficients $c_i$ or $c_j$
tends to zero, the pseudomagnitude tends to the magnitude $m_i$ or
$m_j$. The pseudomagnitude is by construction a reddening free
distance indicator. It can be written as:  
\begin{equation}
pm_{\{i,j\}} = \frac{c_iM_j-c_jM_i}{c_i-c_j}+DM
\end{equation}
where $M_i$ and $M_j$ are absolute magnitudes and $DM$ is the distance
modulus. At this stage, we define the absolute pseudomagnitude
$PM_{\{i,j\}}$ as: 
\begin{equation}
PM_{\{i,j\}} = \frac{c_iM_j-c_jM_i}{c_i-c_j}=pm_{\{i,j\}}-DM
\label{eq:absolute_pseudomagnitude}
\end{equation}
The absolute pseudomagnitude is a reddening free luminosity luminosity
indicator that can be computed very easily. This
requires the knowledge of two magnitudes and a distance. On the other
hand, once the mean absolute pseudomagnitude has been calculated for a
group of stars sharing the same physical properties, the distance
modulus of a star from the same group can be estimated with the
knowledge of just two magnitudes. 
%
%
\section{Absolute pseudomagnitudes of dwarf stars}
\label{sec:absolutepseudomagnitudes}
For our calculations,
we use Eqs.\ref{eq:pseudomagnitude} and~\ref{eq:absolute_pseudomagnitude},
with the second reduction of Hipparcos parallaxes
\citep{2007A&A...474..653V}, the spectral type and the magnitude pairs
(V,J), (V,H) and (V,Ks) provided by SIMBAD. We adopt the interstellar
extinction coefficients determined by \cite{Fitzpatrick99},
thus leading to the following expressions for the pseudomagnitudes:   
%
\begin{eqnarray}
pm_{\{V,J\}} & = & 1.389 \times m_{J}-0.389 \times m_V \nonumber \\
pm_{\{V,H\}} & = & 1.205 \times m_{H}-0.205 \times m_V \nonumber \\
pm_{\{V,Ks\}} & = & 1.136 \times m_{Ks}-0.136 \times m_V
\label{eq:pmvk}
\end{eqnarray}
\subsection{Hipparcos data}
A priori, the absolute pseudomagnitude of a group of stars with the
same spectral type and luminosity class should be constant as a
function of distance. Figure~\ref{fig:1}a 
plots the pseudomagnitude (V,Ks) of Hipparcos class III and V stars
with a spectral type K0 (3747 objects), as a function of their
distance modulus. Figure~\ref{fig:1}b
shows the absolute pseudomagnitude (V,Ks), with the dwarfs lying at the
top and the giants lying at the bottom. For the
same class of stars it is firstly constant, to within the limits
resulting from noise, but beyond a certain distance it then
appears to decrease. 
It is a
mere artifact, due to the fact that below 10\% noise, the inverse of the parallax
begins to be numerically biased.   In this example, 75\% of the
dwarfs and only 26\% of the giants have a parallax noise smaller than
10\%.
\subsection{Practical absolute pseudomagnitude calculation}
\begin{figure}
   \centering
   \includegraphics[angle=90,width=8cm]{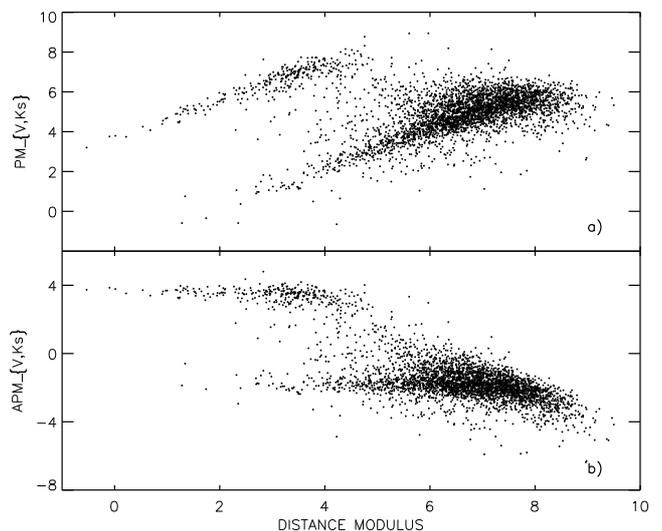}
      \caption{a) (V,Ks) pseudomagnitudes of the 3747
        Hipparcos K0 class III and V stars as a function of their distance
        modulus; b) Absolute (V,Ks) pseudomagnitudes of the same
        stars, with the dwarfs lying at the top and the giants lying
        at the bottom of this figure. The decrease of
        the pseudomagnitudes beyond a certain distance is an artefact
        due to numerical bias at low signal to noise ratio (see
        text). }
   \label{fig:1}
   \end{figure}

In order to calculate the mean absolute pseudomagnitudes of
dwarf stars, we proceed as follows: a) we consider all of the stars in
the Hipparcos catalogue having the same spectral type, with or without
selecting their luminosity class, depending on the possible degree of
confusion; b) we place a limit on the distance of the sample in order
to minimize the influence of the numerical bias\footnote{For example, in the case
  of the K0 stars of figure~\ref{fig:1}\,b, this limit would be around
  $DM=4$
  for dwarves and $DM=7$
  for giants.}; c) since we do not control the
astrophysical biases (see below), we assume that all of the objects 
are statistically equivalent, and adjust the fit of the absolute
pseudomagnitude distribution to one, or even ---in some cases--- to two
Gaussian functions.

This is a difficult operation because the absolute
pseudomagnitude distribution is not always strictly Gaussian.
In practice, stars from the same luminosity class and with the same
spectral type often have stratified luminosities as a function of
their distance. This phenomenon confirms what was already known,
i.e. that for any given spectral type and class of luminosity, there
are hidden sub-classes of stars with distinct physical
properties. Although the absolute pseudomagnitudes would permit a
detailed investigation of these physical properties, for the time
being we do not have sufficient statistical information to implement
such an analysis. This will become possible when the measurements
provided by GAIA~\citep{GAIA} become available.
%
%

\begin{figure}
   \centering
   \includegraphics[angle=90,width=8cm]{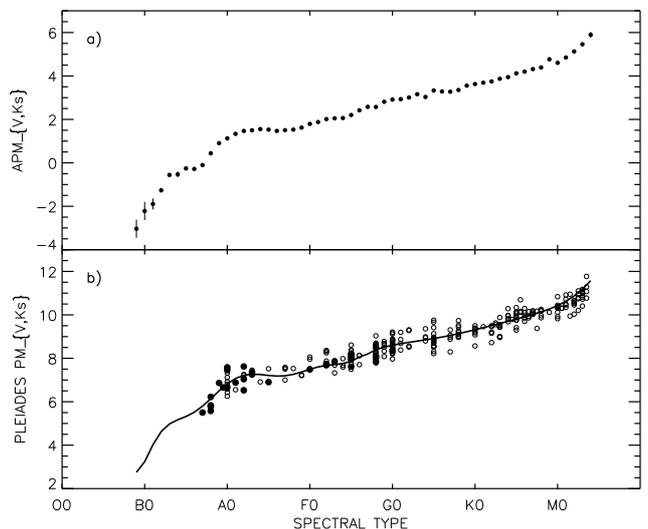}
      \caption{ a) Mean (V,Ks) absolute pseudomagnitudes of field  dwarf stars
        as a function of  
        spectral type, b) Open circles: (V,Ks)
        pseudomagnitudes of 280 Pleiades 
        stars located at less than 0.84 mag (3 times the Gaussian dispersion of
        figure 3) from the Pleiades barycentric
        distance modulus. Hipparcos stars are identified
        with larger filled circles, superimposed our main sequence
        model shifted at the Pleiades distance.} 
         \label{fig:2}
   \end{figure}

Manual calculations were made for each spectral type, and were
repeated several times on various samples of stars. These were based
on the analysis of the pseudomagnitudes of approximately $6000$
dwarf stars, distributed over 56 spectral
sub-types. It corresponds to about 25\% of the Hipparcos stars
identified as dwarfs. 90\% of the selected data have a parallax with
less than 10\% noise, 98\% less than 20\%. 
Figure~\ref{fig:2} shows the mean absolute pseudomagnitudes
(V,Ks) of these dwarf stars as a function of their spectral types,
ranging from O9 to M4.

The median statistical error on the mean absolute
pseudomagnitudes is equal to $0.03$ magnitudes, which corresponds to
an error of $1.5\%$ in terms of distance. For a given group of stars,
the observed dispersions can be 
accounted for by the natural width of the group, which is increased
by the influence of multiplicity, errors of magnitude, distance and
classification. To a lesser extent, they also reflect the star's age
or metallicity. We estimate, to within a factor of 2, that the
systematic error on a correctly characterised single dwarf star is of
the order of $0.05$ magnitude.  

Although pseudomagnitudes have many potential
applications, the most immediate of these is the determination of
the mean distance of a spatially concentrated group of stars, as
for example in the case of stellar clusters and 
galaxies. In the following section we calculate the centroid of the
distance distribution of $360$ stars in the Pleiades cluster, and
whenever possible compare our results with those obtained by other
authors. 
%
\section{Pseudomagnitude distance of the Pleiades}
\label{sec:pleiades}
The Pleiades is one of the most commonly observed young open clusters,
and the properties of its stars provide a \textit{de facto} definition
of the properties of main sequence stars at age zero. Numerous studies
continue to be published regarding the census of this cluster's coeval stars,
and the highest possible accuracy is needed in their distance
determinations in order to test the models of stellar structure and 
evolution. The pseudomagnitude method can be applied to all of the
stars in this cluster, for which the spectral type and at least one
pair of magnitudes is known. It is perfectly adapted to the
calculation of the cluster's mean distance, and could even be
sufficient for the accurate evaluation of the individual distances of
these stars (see section~\ref{vlbi}).
\begin{table}
\tiny
\begin{tabular}{llll}
\hline\T
Refs & Method & N & DM / distance (pc) \B \\
\hline \hline \T
1 & Hipparcos first release & 54 & 5.32 (0.05) / 115.9 (2.7)\\ 
2 & Photometry & 55 & 5.60 (0.04) / 131.8 (2.4) \\
3 & Moving cluster & 65 & 5.58 (0.18) / 130.6 (11.) \\
4 & Ground parallax & 9 & 5.58 (0.12) / 130.6 (7.0) \\
5 & Photometry & 30 & 5.61 (0.03) / 132.4 (1.8) \\
6 & Hipparcos (Makarov) & 54 & 5.55 (0.06) / 129.0 (3.3) \\
7 & Binary & 1 & 5.60 (0.03) / 131.8 (1.8) \\
8 & Binary & 1 & 5.65 (0.03) / 134.9 (1.9) \\
9 & Binary & 1 & 5.60 (0.07) / 131.8 (4.2) \\
10 & HST parallax & 10 & 5.66 (0.06) / 135.5 (3.7) \\
11 & HST parallax & 3 & 5.65 (0.05) / 134.9 (3.1) \\
12 & Binary & 1 & 5.72 (0.05) / 139.3 (3.2) \\
13 & Hipparcos (van Leeuwen) & 54 & 5.40 (0.03) / 120.2 (1.7) \\
14 & VLBI & 5 & 5.67 (0.02) / 136.2 (1.2) \\
15 & Binary & 1 & 5.61 (0.08) / 132.4 (4.9) \\
16 & Photometry & 120 & 5.62 (0.03) / 132.7 (1.8) \\
This work & Pseudomagnitude & 360 & 5.715 (0.018) / 139.0 (1.2) \B\\
\hline
\end{tabular}
\caption{\tiny Measured distances of Pleiades stars, errors are
  between parenthesis. 1: \cite{vanLeeuwen98},
  2: \cite{Pinsonneault1998}, 3: \cite{Narayanan99}, 4: \cite{Gatewood00},
  5: \cite{Stello2001}, 6: \cite{Makarov02}, 7: \cite{Munari04},
  8: \cite{2004Natur.427..326P}, 9: \cite{Zwahlen04}, 10: \cite{Johns-Krull05},
  11: \cite{Soderblom05}, 12: \cite{Southworth05}, 13: \cite{vanLeeuwen09},
  14: \cite{Melis14}, 15: \cite{David16}, 16: \cite{Kim2016}; N:
  target number; DM: distance modulus. 
}
\label{pleiades:distances}
\end{table}
\subsection{On the Pleiades distance controversy }
Whereas an history of distance estimations of the Pleiades cluster can
be found in \cite{An07} and \cite{Melis14},
Table~\ref{pleiades:distances} provides a summary of the measurements
published in the last $20$
years. Various methods have been used. Excluding Hipparcos, the other
direct distance measurements (ground and spaceborne parallaxes,
binaries, VLBI) have relied on the analysis of a total of $\approx30$
stars, and position the Pleiades at a distance between 130 and
139\,pc. The indirect photometric methods were applied on a total of
$\approx 120$ stars and have positioned the cluster at a distance of
132\,pc. In contrast, the mean distance of $54$ Pleiades stars
of spectral types B, A and F by Hipparcos~\citep{vanLeeuwen09} lead to
the controversial distance of $120.2\pm1.7$\,pc, which is indeed
markedly lower (by 10\%) than all other measurements. 
  
It should be recalled that the Pleiades cluster probably contains more
than one thousand stars. When projected onto the sky, it extends over
a distance of the order of 10 to 20\,pc, and it would be reasonable to
assume that the Pleiades has a similar size along its line of sight
when viewed from Earth. Under these conditions, the distances measured
on a few, or even a few tens of objects, with an accuracy much better
than the cluster's expected size, are representative of these objects
distances only. In view of the size of this cluster, it could well be
possible to find star concentrations at distances of the order of
15\,pc from one another. The controversy does not have as much to do
with the so-called distance of the Pleiades cluster\footnote{We
  observe that, given the currently achievable precision on an
  individual star distance and the size of the cluster compared to its
  distance ($\approx10\%$),
  the concept of ``Pleiades distance'' is bound to loose its intended
  meaning.}, as with the mean distance of the 54-odd stars used in the
Hipparcos estimate.

It is difficult to
compare various distance measurements, as they are based on generally
small and generally disjoint samples of stars. The Hipparcos sample
was not used by other independent distance estimations, it was only
reused in new attempts to refine the Hipparcos reduction, first by
\cite{Makarov02} which led to a distance of $129.0\pm3.3$\,pc,
and then by \cite{vanLeeuwen09}, who determined a value of only
$120.2\pm1.7$\,pc.
We note that in view of their uncertainties, these two distance
estimations are only marginally ($2.4\,\sigma$) different.
   \begin{figure}
   \centering
   \includegraphics[angle=90,width=8cm]{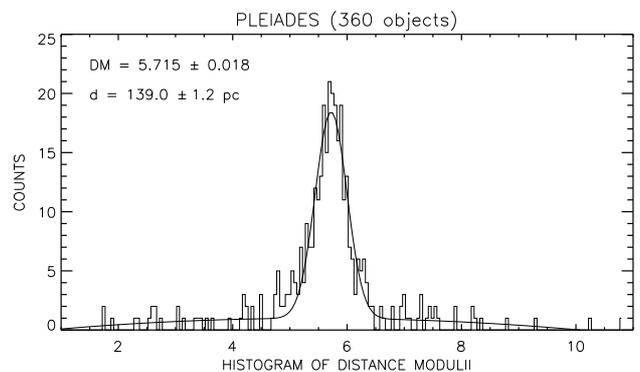}
      \caption{ Distance moduli distribution for 360 Pleiades stars,
        fitted by a Gaussian distribution plus second degree
        polynomial. The Gaussian dispersion (0.28 mag) is dominated by
        spectral classifications errors.} 
         \label{fig:3}
   \end{figure}

Our absolute pseudomagnitude calibration allows us to evaluate the
distance of any sample of stars. In the following section, we
calculate the distance of $360$ Pleiades stars, as well as that of the
Hipparcos sample.
\subsection{Distance of 360 Pleiades stars}
In this section, we assume that Pleiades stars have, at the same
spectral types, the same pseudomagnitudes (V,J), (V,H) and (V,Ks) that 
field dwarfs. Significant differences
occur for cool stars somewhere within the M spectral class. Our sample
of Pleiades stars was obtained from a 
total of $3721$ stars associated with the ``M45'' identifier in the Simbad
database. After filtering (multiplicity, variability, etc.), a total
of $512$ stars remained, of which only $360$
had the required information for the calculation of their distance. As
the Pleiades cluster is very young, in order to increase the size of
our sample, we assumed all of the selected stars to be of luminosity
class V. As the pseudomagnitude is sensitive to the luminosity class,
any non-dwarf star will contribute to the broadening of the distance
distribution, or will get a distance very different to that of the
cluster and will thus be excluded from the analysis. We did not try to
perform filtering for membership, non members will
form a diffuse background that is taken into account in our
statistical modeling.

The adopted distance modulus of each object is the average of the
distance moduli computed from the photometric pairs (V,J), 
(V,H) and (V,Ks), and its error is the dispersion of the three
estimates. 
The (V,Ks) pseudomagnitudes
of our sample, outliers excluded, are shown in Figure~\ref{fig:2}b as
a function of the spectral type. It is not a classical color-magnitude 
diagram. The observed dispersions per spectral type, 0.2 to 0.4 magnitude,
are not imposed by the physics of the cluster but by spectral 
classifications errors, which is probably the limiting noise of our
present approach. We fit the resulting
distance modulus distribution by a Gaussian plus a second degree
polynomial, see Figure~\ref{fig:3}. The centre of the Gaussian
function provides the barycentre of the distance moduli of the $360$
stars studied, i.e. $5.715\pm0.018$, which corresponds to a distance
of $139.0\pm1.2$\,pc.
Although this comparison is somewhat risky, in view of the small
samples used previously, our distance calculation is globally in
agreement with most estimations, but tends to position the cluster at
the high end of measured ``distances''.

What of the stars measured by Hipparcos? We have all of the
information needed to characterise 44 of the 54 stars given in the
list of \cite{Makarov02}. The distribution of their distance moduli
exhibits two maxima, at approximately $5.4$ and $5.7$, a possible
indication of sub-clustering. A gaussian fit of this distribution
leads to a mean distance modulus of $5.66\pm0.06$, i.e. a distance of
$135.5\pm3.7$\,pc, respectively $1.3\sigma$ and $3.8\sigma$
above \cite{Makarov02} and \cite{vanLeeuwen09} estimates. Our result
tends to confirm that on average Hipparcos distances of these stars are
underestimated. Soon we will have the answer on who is right or who is
wrong. But the answer probably will not be as simple as yes or no. 

However, the baby should not be thrown out with the bathwater, since
all of our distance moduli were obtained using absolute
pseudomagnitudes derived from correctly distance-filtered
Hipparcos parallax measurements. The fact that we obtain a barycentric
distance that is compatible (and probably more accurate in terms of
defining the cluster's centroid, as a consequence of the much greater
sample size) with distances measured from the ground, together 
with the fact that we are able to apparently correct the same
controversial Hipparcos measurements, indicates that Hipparcos
parallaxes at large are robust.
\subsection{Distance to the VLBI stars}
\label{vlbi}
Among recent distance measurements, those of \cite{Melis14} determined
by VLBI are the most accurate. As they make it possible to test the
robustness of our pseudomagnitude estimations, we calculate the
distance of $6$ of the $10$ stars scheduled for VLBI observation by
\cite{Melis13} (the 4 others are either not single dwarfs or lacking
spectral type information). Table~\ref{tab:melis_distances} summarises
our predicted distances. For the two stars in common with
\cite{Melis14}, the agreement between VLBI and pseudomagnitude
distances is remarquable, with relative differences of 1\% ($0.5
\sigma$) and 4\% ($1.6\sigma$). 
\section{Conclusion}
Pseudomagnitudes are remarkable distance indicators, since they are
free of interstellar reddening effects. We have calculated the mean
absolute pseudomagnitudes of field dwarfs from O9 to M4, based on the
Hipparcos parallax measurements of approximately 6000 stars, allowing
us to estimate the distance of 360 Pleiades stars. We position the
centroid of these stars at $139.0\pm1.2$\,pc, and we confirm that
the Pleiades stellar distances measured by Hipparcos are on average
underestimated by 10\%. 
\begin{table}
\tiny
\begin{tabular}{llll}
\hline\T
HII & SpT & PMD (pc) & VLBI distance (pc) $^{(1)}$ \B\\
\hline \hline \T
75&G7 &136.2 (3.6)&  \\
253&G1 &143.7 (2.1)&  \\
625&G5 &137.0 (2.4)&138.4 (1.1) \\
1136&G7 &141.0 (3.3)&135.5 (0.6) \\
1883&K2 &139.0 (1.4)& \\
2244&K2 &145.1 (2.1)&\B\\
\hline\B
\end{tabular}
\caption{\tiny Pseusomagnitude distance (PMD) of 6 Pleiades stars of
 the \cite{Melis13} list. (1) \cite{Melis14}}
\label{tab:melis_distances}
\end{table}

ESA's recently launched GAIA mission will make it possible to
accurately determine the fine structure of absolute pseudomagnitudes,
their natural width, and the influence of various parameters such as
age and metallicity. It will be possible to calibrate these very
accurately, in several different optical bands. But already, our initial
results obtained with the Pleiades cluster, together with their
comparison with VLBI measurements, are very encouraging. This
technique is purely observational, direct and simple to implement,
since it needs the knowledge of only the spectral type, two magnitudes
and the corresponding absolute pseudomagnitude.  
\begin{acknowledgements}
   This research has made use of NASA's Astrophysics Data System, of the SIMBAD database
  \citep{2000A&AS..143....9W}, and of the VizieR catalog access
  tool \citep{2000A&AS..143...23O}, CDS, Strasbourg, France. We used the
  TOPCAT tool\footnote{available at
    \url{http://www.starlink.ac.uk/topcat/}}\citep{2005ASPC..347...29T}
 to easily manipulate the star databases used. 
\end{acknowledgements}
\bibliographystyle{aa} 
\bibliography{paper2} 

\onecolumn
\begin{longtab}
\begin{longtable}{lll}
\caption{\label{tab:our_distances} Pseudomagnitude distance (PMD) and
  errors (pc) of 360 Pleiades stars. Names are those returned by CDS
  when searching for 
  ``M45''. The PMDs are the mean of the 3 distances (V,J), (V,H),
  (V,K). The error is the dispersion of those distances. Only the
  subset of 360 stars for which our method is applicable are reported.
}\\ 
\hline\hline
Name&PMD (pc)&Error (pc)\\
\endfirsthead
\caption{continued.}\\
\hline\hline
Name&PMD (pc)&Error (pc)\\
\hline
\endhead
\hline
\endfoot
BD+17 558&179.47&11.01\\
Cl* Melotte 22 MSK 211&144.95&1.01\\
HD 23326&147.74&.54\\
* 22 Tau&114.57&1.61\\
HD 23195&140.02&1.66\\
HD 282960&126.31&.90\\
V* V1084 Tau&114.71&2.52\\
V* V623 Tau&160.84&.03\\
Cl* Melotte 22 HII 1593&142.55&1.99\\
Cl* Melotte 22 DH 507&132.47&1.48\\
V* V1272 Tau&143.68&2.14\\
2MASS J03461174+2437203&151.53&2.01\\
V* V1288 Tau&136.17&3.57\\
Cl* Melotte 22 DH 290&138.06&1.91\\
V* V1046 Tau&119.18&2.65\\
BD+23 521&134.79&.80\\
HD 23568&142.38&.56\\
Cl* Melotte 22 SK 671&115.15&1.00\\
BD+22 521&153.75&.08\\
HD 282965&117.26&3.60\\
V* V540 Tau&157.70&.38\\
Cl* Melotte 22 HII 974&137.64&1.12\\
HD 24087&114.78&2.25\\
V* LT Tau&99.67&2.60\\
V* V1187 Tau&156.35&1.82\\
Cl* Melotte 22 DH 131&116.10&1.08\\
V* V815 Tau&139.99&.88\\
Cl* Melotte 22 MSK 184&141.33&.66\\
BD+26 592&126.75&1.74\\
HD 23312&151.37&1.27\\
V* MS Tau&144.56&1.64\\
HD 23872&139.56&1.74\\
BD+22 574&158.92&1.14\\
* q Tau&92.06&2.07\\
Cl* Melotte 22 DH 525&137.69&.54\\
BD+22 548&126.47&1.71\\
Cl* Melotte 22 SRS 80212&128.48&3.10\\
V* V715 Tau&136.53&.90\\
* 21 Tau&117.65&.95\\
Cl* Melotte 22 SRS 52852&124.27&.91\\
HD 282967&117.87&2.25\\
HD 23464&67.20&1.05\\
HD 23061&150.41&.67\\
V* LV Tau&125.90&.46\\
V* V642 Tau&106.17&1.26\\
HD 282971&149.25&.96\\
HD 23732&131.41&1.08\\
Cl* Melotte 22 SK 775&175.25&.23\\
V* V814 Tau&136.98&.19\\
HD 283046&193.01&2.21\\
Cl* Melotte 22 MSH 175&134.88&1.80\\
HD 283132&134.70&2.24\\
TYC 1799-272-1&144.22&2.83\\
HD 23873&139.10&1.51\\
TYC 1803-1156-1&133.77&1.33\\
Cl* Melotte 22 HII 1110&134.36&1.49\\
2E 857&133.60&4.08\\
BD+23 527&150.35&.08\\
V* V497 Tau&144.71&1.23\\
* 27 Tau&46.63&.54\\
V* V811 Tau&137.04&2.42\\
Cl* Melotte 22 DH 212&128.57&1.96\\
HD 24194&126.86&.60\\
V* V641 Tau&143.41&2.07\\
Cl* Melotte 22 DH 293&157.91&1.91\\
BD+21 504&120.15&2.06\\
Cl* Melotte 22 DH 267&112.64&2.81\\
V* V1274 Tau&47.75&.69\\
2MASS J03441466+2406065&91.95&3.08\\
HD 23763&100.51&1.10\\
HD 282975&98.92&.81\\
Cl* Melotte 22 MSK 74&140.34&4.94\\
Cl* Melotte 22 SK 40&162.67&.75\\
V* OS Tau&115.60&3.38\\
Cl* Melotte 22 DH 108&123.33&2.93\\
Cl* Melotte 22 DH 184&141.75&1.00\\
V* V1010 Tau&133.73&.54\\
TYC 1803-1351-1&98.47&.85\\
V* V1228 Tau&117.20&.61\\
V* V644 Tau&139.39&2.44\\
HD 23608&113.83&1.24\\
V* V378 Tau&160.17&.64\\
Cl* Melotte 22 SK 488&133.65&3.45\\
Cl* Melotte 22 HHJ 437&207.91&2.11\\
HD 23924&161.74&2.23\\
SAO 76387&185.78&1.88\\
Cl* Melotte 22 DH 143&134.44&.87\\
V* V476 Tau&97.93&.16\\
2MASS J03493653+2417460&179.22&1.55\\
Cl* Melotte 22 DH 436&146.34&.51\\
Cl* Melotte 22 DH 562&127.81&3.78\\
HD 23387&108.94&1.44\\
V* V446 Tau&143.41&2.00\\
HD 283420&110.87&.51\\
BD+22 553&169.93&2.09\\
HD 282958&126.44&1.88\\
Cl* Melotte 22 SK 754&150.36&3.04\\
CCDM J03481+2409AB&111.98&1.84\\
V* V703 Tau&133.83&.41\\
* 16 Tau&130.34&1.47\\
V* V1065 Tau&140.96&3.31\\
Cl* Melotte 22 SSHJ G315&165.54&1.70\\
V* V664 Tau&145.15&3.11\\
Cl* Melotte 22 HII 102&122.18&1.24\\
HD 24463&114.36&1.94\\
HD 282973&117.76&3.80\\
V* V727 Tau&96.25&.99\\
V* V1224 Tau&148.57&5.47\\
BD+23 472&150.45&1.28\\
BD+22 624&143.38&.79\\
HD 23352&164.74&3.79\\
Cl* Melotte 22 DH 153&93.01&2.47\\
V* V855 Tau&114.57&3.98\\
Cl* Melotte 22 DH 875&155.64&1.71\\
HD 23327&128.01&.71\\
* eta Tau&41.17&.44\\
Cl* Melotte 22 DH 349&148.80&.71\\
BD+22 552&167.99&.44\\
HD 23975&113.28&1.96\\
Cl* Melotte 22 DH 603&133.62&1.12\\
Cl* Melotte 22 DH 734&177.77&1.85\\
HD 23886&154.19&1.30\\
HD 22444&89.73&1.89\\
V* V700 Tau&129.73&.45\\
Cl* Melotte 22 SRS 68435&184.05&.18\\
BD+25 555&148.92&2.95\\
V* V647 Tau&160.26&2.34\\
Cl* Melotte 22 DH 486&139.59&2.71\\
Cl* Melotte 22 SK 709&138.63&1.28\\
HD 23935&123.82&3.02\\
Cl* Melotte 22 MSH 82&161.58&1.22\\
V* V810 Tau&176.53&8.31\\
BD+23 551&140.10&2.78\\
V* V650 Tau&142.39&1.35\\
V* V652 Tau&103.63&.37\\
V* V966 Tau&141.91&1.82\\
TYC 1799-102-1&154.68&1.44\\
V* V813 Tau&195.53&6.05\\
* 17 Tau&69.10&1.15\\
V* LR Tau&33.04&.61\\
HD 23514&146.05&2.29\\
V* V812 Tau&131.48&.75\\
Cl* Melotte 22 LLP 15&140.66&1.43\\
V* V1041 Tau&131.69&2.10\\
UCAC2 40300217&109.71&1.29\\
Cl* Melotte 22 DH 421&106.65&.82\\
V* V1045 Tau&162.03&2.77\\
HD 23351&133.79&1.26\\
Cl* Melotte 22 DH 417&149.93&.80\\
Cl* Melotte 22 DH 462&140.35&.28\\
BD+20 672&121.96&1.68\\
V* V1283 Tau&128.04&2.36\\
V* V1210 Tau&139.54&1.26\\
Cl* Melotte 22 DH 456&162.22&2.37\\
* 20 Tau&52.64&1.86\\
V* PR Tau&149.08&1.09\\
Cl* Melotte 22 DH 271&133.39&2.75\\
Cl* Melotte 22 MSK 44&161.22&3.76\\
V* V643 Tau&110.74&.65\\
NAME 1RXS J034412.1+240200SE&135.08&2.99\\
V* OU Tau&151.78&3.78\\
V* V1170 Tau&133.91&2.06\\
V* KO Tau&118.00&5.35\\
HD 23511&148.60&3.30\\
V* V1175 Tau&148.48&3.71\\
Cl* Melotte 22 K 78&135.25&3.82\\
V* V382 Tau&117.31&1.33\\
V* V534 Tau&130.10&1.40\\
V* V660 Tau&139.00&1.43\\
Cl* Melotte 22 LLP 28&103.66&2.49\\
V* V1090 Tau&153.17&3.79\\
V* V371 Tau&94.09&1.18\\
V* V535 Tau&145.83&.36\\
V* V1169 Tau&153.65&2.73\\
V* V1171 Tau&310.83&11.59\\
V* V969 Tau&90.48&1.03\\
Cl* Melotte 22 HII 2209&151.73&3.20\\
HD 282954&133.28&.95\\
HD 23513&148.86&1.06\\
HR 1183&142.46&.91\\
V* V1282 Tau&101.59&.96\\
HD 23489&130.18&2.37\\
HD 24665&116.06&1.25\\
BD+27 545&159.50&1.79\\
HD 23512&142.27&1.43\\
HD 23791&157.58&4.36\\
V* V1176 Tau&100.40&1.12\\
V* V1193 Tau&143.70&.66\\
HD 23584&139.06&2.88\\
HD 23598&134.56&1.77\\
Cl* Melotte 22 DH 730&149.09&2.56\\
V* V963 Tau&131.11&1.08\\
V* V1173 Tau&141.24&1.39\\
HD 23912&146.43&.76\\
HD 23158&148.63&.53\\
HD 23361&147.82&1.48\\
HD 23409&143.08&1.41\\
V* V1172 Tau&143.41&2.13\\
V* V816 Tau&152.73&4.91\\
V* V545 Tau&151.08&1.32\\
V* V844 Tau&163.42&2.94\\
HD 23479&125.96&.22\\
HD 23733&131.23&.71\\
HD 23632&126.62&4.35\\
* 18 Tau&119.70&1.26\\
HD 23863&157.47&.78\\
HD 23778&127.42&.27\\
V* V785 Tau&142.26&.94\\
V* V1174 Tau&189.61&3.19\\
HD 282952&159.92&6.83\\
BD+23 513&128.94&2.07\\
HD 24076&102.17&.69\\
HD 23269&127.53&3.76\\
* 24 Tau&104.48&1.15\\
V* II Tau&29.56&.22\\
BD+19 587&143.25&.63\\
* 28 Tau&82.99&3.27\\
HD 23610&179.87&1.89\\
HD 23948&171.39&1.86\\
HD 24132&138.60&1.68\\
BD+21 508&153.17&2.71\\
V* V370 Tau&159.63&.39\\
V* V518 Tau&184.17&3.43\\
V* V677 Tau&136.39&1.51\\
* 23 Tau&78.99&.69\\
HD 23375&136.76&2.21\\
HD 23631&156.43&1.05\\
Cl* Melotte 22 DH 304&150.95&1.83\\
Cl* Melotte 22 MSH 74&146.85&1.10\\
HR 1172&106.54&.82\\
Cl* Melotte 22 MSK 140&137.55&3.96\\
GJ 3219 A&29.19&.21\\
V* V452 Tau&87.67&1.61\\
BD+20 628&213.30&1.91\\
BD+23 514&143.40&3.25\\
HD 283117&129.62&1.59\\
V* V539 Tau&91.22&.18\\
HD 283031&739.44&19.91\\
BD+24 501&326.31&1.98\\
Wolf 1260&89.06&2.21\\
V* CL Ari&176.87&2.72\\
V* V1085 Tau&182.90&.50\\
Cl* Melotte 22 SK 792&325.49&1.26\\
HD 282926&582.29&16.02\\
BD+24 470&453.13&9.08\\
HD 283222&110.23&1.20\\
V* V1227 Tau&10.51&.72\\
V* V613 Tau&123.89&1.72\\
V* V372 Tau&101.83&1.31\\
HD 23157&118.59&.60\\
TYC 1807-1756-1&183.99&3.55\\
BD+24 456&436.57&6.13\\
HD 22693&189.20&5.42\\
HD 283079&177.62&1.76\\
TYC 1805-572-1&248.78&1.59\\
HD 282998&142.82&1.77\\
Cl* Melotte 22 SK 646&164.77&.89\\
HD 24105&73.22&3.48\\
V* V349 Tau&112.88&2.23\\
V* V638 Tau&142.50&.79\\
BD+26 580&102.94&.85\\
Cl* Melotte 22 WCZ 141&1119.62&18.23\\
HD 283044&42.57&.80\\
V* V532 Tau&143.20&2.38\\
V* SZ Ari&304.14&10.70\\
HD 283058&62.35&1.11\\
V* LO Tau&118.57&2.20\\
V* V468 Tau&179.05&2.66\\
V* PP Tau&143.72&2.30\\
GJ 3227&24.03&.14\\
V* V338 Tau&124.47&1.43\\
V* V377 Tau&137.61&1.27\\
Cl* Melotte 22 DH 368&170.68&4.08\\
HD 282942&49.30&.75\\
BD+25 604&125.24&1.53\\
V* CG Ari&80.46&1.54\\
GJ 3240&22.47&.33\\
V* V561 Tau&137.48&2.05\\
V* EQ Tau&178.99&1.32\\
V* QS Tau&1383.72&90.87\\
StKM 1-406b&53.58&.32\\
V* V361 Tau&131.28&1.51\\
LP 355-27&33.41&.75\\
HD 23431&260.15&1.28\\
BD+24 479&340.82&7.25\\
V* V358 Tau&133.71&2.42\\
V* V366 Tau&101.41&1.12\\
BD+18 541&121.01&1.33\\
HD 283014&437.72&3.74\\
BD+19 589&107.11&4.69\\
BD+26 586&314.70&1.70\\
BD+26 553&283.42&3.47\\
BD+17 637&228.69&5.36\\
HD 23410&130.63&4.62\\
HD 23289&142.45&1.22\\
V* V679 Tau&88.24&1.01\\
V* V502 Tau&144.40&.95\\
V* V357 Tau&149.65&2.15\\
BD+20 549&154.47&1.80\\
HD 283139&125.47&1.44\\
BD+23 433&285.71&2.96\\
V* FL Tau&113.78&1.62\\
HD 285234&145.61&1.23\\
Cl* Melotte 22 DH 504&107.90&1.18\\
HD 282972&154.18&6.75\\
V* V1229 Tau&122.60&.45\\
TYC 1787-384-1&247.45&4.40\\
HD 283038&230.53&2.11\\
V* V380 Tau&126.91&1.41\\
V* V470 Tau&154.77&2.24\\
V* V354 Tau&128.81&2.72\\
BD+20 626&109.33&2.22\\
BPM 85549&68.12&.56\\
2MASS J03235551+2339273&65.16&3.11\\
BD+16 455&304.16&1.27\\
BD+19 594&150.76&2.43\\
V* V739 Tau&130.51&3.61\\
V* V376 Tau&154.42&3.50\\
BD+19 607p&179.56&2.80\\
BD+25 592&246.77&4.79\\
2MASS J03273245+2554003&73.26&1.41\\
BD+22 512&32.64&.45\\
HD 283032&423.77&8.11\\
HD 24344&378.83&1.53\\
HD 283055&161.93&2.79\\
V* V1286 Tau&126.09&.56\\
NAME 1RXS J034412.1+240200NW&196.43&15.41\\
HD 285243&93.75&.88\\
HD 283006&141.52&1.05\\
TYC 1798-1002-1&125.91&2.30\\
BD+25 539&170.07&1.61\\
LH98 95&119.66&1.10\\
V* CU Tau&260.27&6.18\\
BD+20 565&253.31&2.43\\
Cl* Melotte 22 MSK 100&117.43&5.48\\
V* V399 Tau&129.12&.78\\
GJ 3239&35.30&.73\\
StKM 1-417&89.95&1.17\\
2MASS J03181744+1824202&72.59&1.94\\
2MASS J03414386+1824061&39.48&.55\\
BD+25 610&129.16&2.21\\
BD+20 594&128.88&.58\\
HD 282955&383.91&3.69\\
HD 283036&282.28&6.21\\
GJ 3225&32.09&.82\\
BD+22 468&181.70&1.81\\
HD 24355&303.59&2.30\\
GJ 140 C&21.81&.23\\
BD+25 572&68.08&1.86\\
TYC 1805-890-1&155.43&.13\\
HD 22139&137.84&1.69\\
V* QX Tau&123.85&.51\\
V* V343 Tau&146.99&2.75\\
HD 24088&214.86&3.73\\
BD+23 538B&93.64&2.68\\
2MASS J03164389+1923041&177.42&5.74\\
V* CK Ari&33.27&.16\\
HD 282928&246.31&.86\\
HD 282990&150.14&2.82\\
HD 23964C&113.88&.93\\
\end{longtable}
\end{longtab}

\end{document}